\documentclass[useAMS]{mn2e}
\title[Metallicity, planetary formation and migration]
{Metallicity, planetary formation and migration}

\author[M. Livio and J.~E. Pringle]{Mario Livio$^1$ and J.~E.\ Pringle$^1$\thanks{Permanent address: Institute of Astronomy, Madingley Road, Cambridge, CB3 0HA, U.K.}\\
$^1$Space Telescope Science Institute, 3700 San Martin Drive, Baltimore, MD 21218; email: mlivio@stsci.edu}

\begin{document}

\pagerange{\pageref{firstpage}--\pageref{lastpage}} \pubyear{2003}

\maketitle

\label{firstpage}

\begin{abstract}
Recent observations show a clear correlation between the 
probability of hosting a planet and the metallicity of the parent 
star. Since radial velocity surveys are biased, however, towards 
detecting planets with short orbital periods, the 
probability-metallicity correlation could merely reflect a dependence 
of migration rates on metallicity. We investigated the possibility, but 
find no basis to suggest that the migration process is sensitive to the 
metallicity. The indication is, therefore, that a higher metallicity 
results in a higher probability for planet \emph{formation}.
\end{abstract}

\section{Introduction}

Planet searches, mostly via radial velocity surveys of nearby stars,
have so far yielded some 110 planets
(e.g.\ Schneider 2003).  The stars that are
harboring planets have been known for some time to be relatively
metal-rich (e.g.\ Gonzalez 1998, 1999; Gonzalez, Wallerstein \& Saar
1999; Reid 2002). While some of the early studies suffered from clear 
selection effects, since the planet hunters tended to concentrate on 
metal-rich stars, more recent studies tend to be merely brightness 
limited.  In particular, Fischer \& Valenti (2003) examined 754
stars in the solar neighborhood and found that the probability of 
hosting a planet rises from
about 5~per cent when the iron abundance is $\sim\!1/3$ that of the
sun, to about 20~per cent, when the iron abundance is $\sim$3~times
that of the sun.  We will generally assume that the metal content of 
the stellar host reflects the metallicity of the planets/protoplanetary 
disc (although see Saumon et~al.\ 1996). Suggestions have been made that 
the observed metallicity `excess' may reflect `pollution' of the star's 
convective envelope by the accretion of planets (e.g.\ Laughlin \& Adams 
1997; Conzalez 1998). If that were true, however, one would expect stars 
with shallow convection zones (e.g.\ F~stars) to show correspondingly 
higher metallicity enhancements. Such an effect is not observed (Fisher 
\& Valenti 2003). 

Broadly speaking, there are two scenarios for planet
formation. One scenario involves a multi-stage process, in which dust
grains coagulate to form rocks, then planetesimals, and the latter
coalesce to form planetary embryos that accrete the gaseous envelopes
(e.g.\ Goldreich \& Ward 1973; Pollack et~al.\ 1996; Lissauer 2001). In the
other scenario, giant planets form from the gravitationally collapsing
regions of unstable protoplanetary discs (e.g.\ Boss 2001; Mayer et~al.\ 
2002; Rice et~al.\ 2003).  On the face of it, the observed correlation 
between the metallicity of a star and its probability of hosting 
planets can be taken as supporting the multi-stage formation process. 
However, it should be realized that all the observed extrasolar planets 
(around main sequence stars) are believed to have migrated inward from 
larger radii (e.g.\ Lin, Bodenheimer \& Richardson 1996; Armitage et~al.\ 
2002; Trilling, Lunine \& Benz 2002), because planet formation at the
observed radii appears to be difficult (e.g.\ Bodenheimer, Hubickyj \& 
Lissauer 2000; Mayer et~al.\ 2002). This raises the possibility, in 
principle, that the higher metallicity is required for the migration, 
rather than for the formation process (Sigurdsson et~al.\ 2003).  In the 
present Letter we examine this possibility.

\section{Migration and metallicity}

In order for the observed metallicity-planet frequency correlation to
be an effect of migration alone, we need to explore the possibility
that a low metallicity somehow inhibits migration. Mechanisms of
migration that involve planet-planet scattering (e.g.\ Rasio \& Ford
1996; Papaloizou \& Terquem 2001) do not depend on the metallicity. The
scattering of planetesimals by massive planets (e.g.\ Murray et~al.\ 
1998) does depend on the metallicity, but only if one requires the
planetesimals to \textit{form} via the multi-stage process.  The only
mechanism we need to study, therefore, is migration due to 
gravitational interaction with a gaseous, viscous, protoplanetary
disc.  The process we are interested in is ``Type~II" migration, in
which the planet opens a gap in the disc, and subsequently follows its
viscous evolution.  We take the view that an observed Jupiter-sized
planet is essentially the \emph{last one} to have formed as the migration 
came to a halt (Armitage et~al.\ 2002; Trilling, Lunine \& Benz, 2003). 
In this case, assuming that Jupiter-sized planets have to form close to, or
beyond, a radius of $\sim$5~AU, (e.g.\ Mayer et~al.\ 2002; Boss 1995) we are
interested in the properties of remnant discs at a time when the 
remaining mass of the disc was a few Jupiter masses at a radius of 
$\sim$5~AU. If the disc is much more massive than this, then the 
subsequent evolution of the disc will sweep the planet into the star. 

There are two obvious ways, in principle, in which migration can be
significantly affected: (i)~If the viscosity in the disc is reduced to
extremely low values due to the MHD turbulence dying away, and (ii)~if
migration becomes significantly slower due to global changes in the
disc structure. In what follows, we examine each of these
possibilities in turn.

\subsection{Dead zones}

The viscosity in accretion discs most probably originates from the
magneto-rotational instability (MRI; Balbus \& Hawley 1991). Numerical
simulations have shown that the MRI, and the concomitant MHD
turbulence are suppressed when the magnetic Reynolds number is
smaller than some critical value (Gammie 1996; Gammie \& Menou 1998;
Fleming, Stone \& Hawley 2000).  This leads in some parts of the disc 
to a layered disc structure, 
in which the gas near the disc mid-plane is cold, shielded from ionizing 
radiation and non viscous.  Accretion then occurs only in a very thin 
surface layer that is ionized by cosmic rays.
 
In a shear flow, the magnetic Reynolds number is defined as
\[ 
Rm = L V/\eta~~, 
\]
where $L$ and $V$ are typical length and velocity scales,
respectively, and $\eta$ is the resistivity. Adopting $L \sim H$ (the
disc half-thickness), $V \sim c_s$ (the speed of sound), as is
appropriate for the simulations that make use of the shearing box
approximation (e.g.\ Fleming et~al.\ 2000), the requirement 
$Rm \la Rm_\mathrm{crit}$ corresponds to
balancing the growth rate of the MRI by Ohmic diffusion. When no
externally imposed vertical magnetic field is present, the critical value 
is of order $Rm_\mathrm{crit} \sim 2000$ (Hawley, Gammie \& Balbus, 1996).
When one allows for an external vertical uniform field, $Rm_\mathrm{crit}
\sim 100$ (Fleming et~al.\ 2000).  The decay of MHD angular momentum
transport results in a ``dead zone," in which the viscosity is very
low, and migration might be expected either to stop or to be slowed down 
significantly. Of course, if the dead zone does not significantly change 
the rate of migration, then the effect of metallicity on the dead zone 
is not relevant to our current considerations. Since the Reynolds number
depends on the resistivity, which, in turn, depends on the electron
fraction, a change in the metallicity could lead to changes in the
formation of a dead zone that could lead to changes in the properties
of migration.  The resistivity is given by (e.g.\ Matsumura \& Pudritz
2003)
\[
\eta = 234\ T^{1/2}/x_e\  \mathrm{cm^2~s}^{-1}~~, 
\]
where $T$ is the mid-plane disc temperature, and $x_e = n_e/n$ is the electron
fraction ($n_e$ and $n$ are the number densities of electrons and
neutral atoms respectively). Thus, at fixed $x_e$, since $H
\propto c_s/\Omega$, where $\Omega \propto R^{-3/2}$ is the angular
velocity of the disc, we have that
\[
Rm \propto T^{1/2} R^{3/2}~~.
\]

Matsumara \& Pudritz (2003) consider the radiative, hydrostatic disc
model developed by Chiang \& Goldreich (1997), and Chiang et~al.\ 
(2001). This model has an assumed surface density of the form
\[
\Sigma = \Sigma_o (R/\mathrm{AU})^{-3/2}~~,
\]
with a fiducial value of $\Sigma_o = 10^3$~g~cm$^{-2}$ which gives a
disc mass of a few Jupiter masses within about 5~AU. The flared,
radiative equilibrium disc of Chiang \& Goldreich (1997) has $T
\propto R^{-3/7}$ for $R < 84$~AU. Assuming a state of ionization
balance (where the ionization can be caused by either the central star
or cosmic rays), and using the equation of the electron fraction given
by Oppenheimer \& Dalgarno (1974), Matsumura \& Pudritz (2003) solved
for the radius of the dead zone in the two limiting cases of no metals
at all, $x_m =0$, and essentially all metals ($x_m \gg x_e$), where
$x_m$ is metal fraction. Assuming a critical Reynolds number
appropriate for a disc with a superimposed poloidal field (which they
take to correspond to a critical Reynolds number $Rm_\mathrm{crit}
=1$--10), they found that even at these extremes, the radius of the dead
zone differed by at most a factor $\sim\!2$ ($\sim\!1$~AU in the metal
dominant case, and $\sim$2~AU in the no-metals case).

For the fiducial disc we see that $Rm \propto R^{9/7}$.  Thus, if one
increases the critical Reynolds number by a factor of $\sim\!20$--100, 
as is inferred from simulations with no imposed poloidal field,
and which is more likely to be the case for a protoplanetary disc at
$R \sim 5$~AU, then the radius of the dead zone becomes 
\textit{independent} of the metallicity, and is determined solely by 
the surface density ($\Sigma \sim 100$~g~cm$^{-2}$) that allows for
ionization by cosmic rays (Gammie 1996). In the disc model used by
Matsumura \& Pudritz (taken from Chiang et~al.\ 2001), this corresponds
to a dead zone radius of $\sim\!5$~AU.  We therefore find that, if one
just considers the effect of dead zones on the disc structure, and
hence on the migration rate, then the migration rate is unaffected,
even if the mass fraction in metals is changed from $Z=1$ to $Z=0$.

\subsection{Disk structure and metallicity}
                              
We now consider whether a change in metallicity leads to a sufficient
change in disc structure that the disc inflow timescale, $\tau_\nu$,
and hence the planet migration rate, changes significantly. Since the
disc models of Chiang \& Goldreich (1997) are static, they are not
suitable for this purpose. Ideally, one would like to consider a
time-dependent model of a proto-planetary disc when its mass has
reached the relevant value (of a few Jupiter masses at $\sim$5~AU). 
However, for simplicity and to get some
indication of the sensitivity of the disc structure to metallicity, we
use the disc models of Bell et~al.\ (1997), who consider the detailed
disc structure for a variety of values of steady accretion rates. In
these models no account is taken of the possibility of ``dead zones,''
and the viscous process is parameterized using the usual
$\alpha$-prescription (Shakura \& Sunyaev, 1973). Recent calculations
by Fleming \& Stone (2003) indicate that this might in fact be a
reasonable approach. Their model with an accretion rate of $\dot{M} =
10^{-8}$~M$_{\odot}$~y$^{-1}$ has a surface density of $\Sigma \approx
140$~g~cm$^{-2}$ at a radius of 5~AU. Thus the mass of the disc at
this radius is of order a few Jupiter masses, as required. What is
relevant for our purposes is the viscous inflow timescale given by
(Pringle 1981)
\[
\tau_\nu \sim R^2/\nu~~,
\]
where $\nu \approx \alpha c_s^2/\Omega$ is the viscosity. The Bell 
et~al.\ (1997) models assume that the dimensionless viscosity parameter 
is $\alpha = 0.01$. Thus at a given radius, $\tau_\nu \propto T^{-1}$, the
mid-plane disc temperature. In the same model the (optically thick) disc 
temperature at $R \sim 5$~AU is $T \approx 30$~K. In this temperature range, 
Bell et~al.\ (1997) find, using Henning \& Stognienko (1996), that the 
opacity is approximately of the form $\kappa = \kappa_o T^{2.1}$, and 
that $T \propto \kappa_o^{0.34}$. Since at these temperatures the
opacity is predominantly due to dust (Pollack et~al.\ 1994; Henning \&
Stognienko 1996), and although there may be complications due to the
detailed structure of the dust and the associated chemistry, it seems
reasonable as a first approximation to assume that $\kappa_o$, is
proportional to the metallicity, i.e.\ $\kappa_o \propto Z$. In this
case a change in metallicity by a factor of $\sim\!10$ leads to a
corresponding change in the inflow timescale, and hence migration 
timescale by a factor of $\sim\!2$.

\section{Discussion and conclusions}

The observed correlation between the metallicity of a star and its
probability for hosting a planet seems to indicate that metallicity is an 
important factor in planet formation. However, the sample of detected 
extrasolar planets suffers from obvious selection effects (e.g.\ Zucker 
\& Mazeh 2001; Tabachnik \& Tremaine 2002; Lineweaver \& Grether 2002;
Armitage et~al.\ 2002). In particular, radial velocity techniques have almost
no sensitivity to planets with orbital periods longer than the duration of
the survey. Consequently, if the migration process can be inhibited in the
low metallicity systems, this would clearly produce a correlation of the type 
that has been observed. 

In the present work, we have therefore examined 
potential effects of low-metallicity on migration. By considering the basic 
physical processes involved, we have shown that in low-metallicity systems, 
migration could be slowed down, but at most by a factor $\sim\!2$ (in terms 
of timescale). Such a modest sensitivity is not expected to produce the 
observed rise in probability for hosting a planet (Armitage et~al.\ 2002 
have considered models in which the viscous timescale was changed by similar 
factors, with no significant consequences for the resultant orbital distribution).
Therefore, our tentative conclusion is that the metallicity observations do 
support the idea that a higher metallicity is associated with a higher rate of 
\textit{formation} of planets rather than with migration. Our conclusion is 
further supported by the fact that there is no distinguishable difference in 
metallicity between the stars hosting planets with semi-major axes longer than 
the observed median, and those with shorter ones (Fischer \& Valenti 2003). 
To make this conclusion ironclad would require a detailed, time-dependent, 
migration computation, that would include a full treatment of the thermal 
structure of the disc. 

It is interesting to note that the non-detection of 
transiting planets in the globular cluster 47~Tuc (Gilliland et~al.\ 2000) 
is entirely consistent with the dependence of hosting a planet on metallicity. 
Originally, Gilliland et~al.\ estimated an expected number of 17 detections 
(from monitoring 34,000 stars). If one takes account, however, of the fact (Marcy 
et~al.\ 2003) that only 0.6\% of the F, G, K, M~stars in the solar neighborhood have 
``hot Jupiters" (giant planets with orbital periods $P<5$~days; 
Gilliland et~al.\ assumed 1\%), and the dependence of probability on 
metallicity, the number of expected detections is reduced to $\sim\!2$. 

If the formation of planets indeed requires metals, one needs to 
explain the recent observation of a planet in the globular cluster M4 
(Sigurdsson et~al.\ 2003). That cluster has a metallicity that is only 5\% that
of the sun, and therefore the formation of planets in it would be expected to be 
suppressed. It is important to note, however, that even in such a low metallicity 
environment, there are circumstances where dust and metals are abundant. One 
example is supernovae (note that the planet in M4 is a companion to the pulsar
B1620$-$26), which can produce copious amounts of dust (2--4 solar masses
in the case of Cas~A; Dunne et~al.\ 2003). Another dust-rich environment is 
around asymptotic giant branch stars. In fact, the white dwarf companion to 
the pulsar B1620$-$26 had to evolve through such a phase. It is not impossible 
that the circumbinary planet was in fact formed during that phase (and not 
around a main sequence star, as suggested by Sigurdsson et al.), and it was 
later pushed via interactions into a non-coplanar orbit. Finally, planets 
around pulsars (or white dwarfs) can form from the disruption of white dwarf 
companions (Livio, Pringle \& Saffer 1992; Podsiadlowski et~al.\ 1991). 

\section*{Acknowledgements}
We are grateful to Jeff Valenti and Ron Gilliland for helpful discussions. JEP 
thanks STScI for hospitality and support from the Visitors' Program. ML 
acknowledges support from John Templeton Foundation Grant 938-COS191. We 
acknowledge useful comments from an anonymous referee.

\label{lastpage}

\end{document}